\documentclass[article]{elsarticle}

\usepackage{lineno,hyperref}
\usepackage{amsmath}
\usepackage{tabularx}

\journal{arXiv}

\newcommand{\rmii}[1]{{\mbox{\tiny\rm{#1}}}}
\newcommand{\beq}{\begin{equation}}
\newcommand{\eeq}{\end{equation}}
\newcommand{\bea}{\begin{align}}
\newcommand{\eea}{\end{align}}
\newcommand{\beas}{\begin{align*}}
\newcommand{\eeas}{\end{align*}}

\newcommand{\eps}{{\varepsilon}}
\newcommand{\mD}{m_\rmii{D}}

\newcommand{\vE}{\vec E}
\newcommand{\rh}{\hat r}
\newcommand{\vr}{\vec r}
\newcommand{\vp}{\vec p}

\newcommand{\Tint}[1]{{\hbox{$\sum$}\!\!\!\!\!\!\!\int\,}_{\!\!\!\!\raise-0.9ex\hbox{$\scriptstyle{#1}$}}}
\renewcommand{\Re}{\rm Re}
\renewcommand{\Im}{\rm Im}

\begin{document}

\begin{frontmatter}

\title{A gauge invariant Debye mass and the complex heavy-quark potential}

\author[YBA]{Yannis Burnier}
\author[ARA]{Alexander Rothkopf}

\address[YBA]{Institute of Theoretical Physics, EPFL, CH-1015 Lausanne, Switzerland}
\address[ARA]{Institute for Theoretical Physics, Universit\"at Heidelberg, Philosophenweg 12,  D-69120 Heidelberg, Germany}

\begin{abstract}
Following the original idea of Debye, we define and extract a gauge-invariant screening mass from the complex static in-medium heavy-quark potential $V_{Q\bar{Q}}$, recently obtained from lattice QCD. To this end we derive a field theoretically motivated analytic formula that faithfully reproduces both the screened real- as well as the imaginary part of the lattice potential with a single temperature dependent fit parameter $m_D(T)$. Using values of the real part of $V_{Q\bar{Q}}$ in a gluonic medium, we obtain Debye masses compatible with predictions from HTL perturbation theory.
\end{abstract}


\end{frontmatter}

\section{Introduction}
The concept of a screening mass helps us to intuitively understand the intricate interplay of interactions that take place, once a test particle is inserted into a medium of charge carriers. Originally Debye and H\"uckel \cite{DebyHueckel} investigated the behavior of ions in electrolyte solutions and found that their interactions could be understood by an exponential suppression of the vacuum Coulomb potential due to the presence of classical Boltzmann distributed charges. This in-medium modification, translated into the modern language of quantum field theory, amounts to a medium induced dressing of the mediating gauge bosons, bestowing them with an otherwise absent mass. In the context of perturbative quantum electro-dynamics this mechanism of a thermal mass generation is well understood, as arising from the zero momentum limit of the gauge invariant longitudinal gluon self energies. In perturbative QCD, only the leading order and the logarithm at next to leading order (NLO) of the Debye mass can be computed, the constant at NLO already receives non-perturbative contributions \cite{Arnold:1995bh}. On the lattice, its definition encounters another difficulty, unlike in QED where the Debye mass can be extracted from the electric field correlator, in QCD the electric field itself is not gauge invariant. Several approaches were proposed to circumvent this problem using e.g. effective theories obtained by dimensional reduction \cite{Hart:2000ha}, spatial correlation functions of gauge invariant meson correlators \cite{DeTar:1987ar} or the behavior of the color singlet free energies \cite{Nadkarni:1986cz,Maezawa:2007fc,Digal:2005ht}. 

Here we return to the original idea of Debye and identify a physical observable, the static heavy-quark potential $V(r)$ between a color singlet configuration of a quark and anti-quark, to non-perturbatively define a screening mass for QCD. In the vacuum the potential exhibits both a perturbative Coulombic, as well as a non-perturbative string-like behavior \cite{Koma:2006si}. Both of these features will receive in-medium modification, as can e.g. be seen in a recent lattice QCD determination \cite{Burnier:2014ssa}. An additional complication arises from the fact that the potential is in general a complex quantity \cite{Laine:2007qy}, due to the presence of scattering of light medium degrees of freedom with the color string spanning in between the heavy quark and anti-quark. A meaningful description of the relevant physics must therefore necessarily capture both the effects of screening of the real part of the potential ${\Re}V(r)$ as well as Landau-damping related to ${\Im}V(r)$. Our strategy hence is to find a field-theory motivated parametrization of the  potential that depends only on a single temperature dependent parameter $m_D(T)$, which we will be able to identify with the Debye screening mass.

In the literature two contributions along this path can be found. On the one hand, Ref.~\cite{Digal:2005ht} proposed an analytic function for the real part of a medium-modified Cornell-type potential by working within the fully classical setup of a test charge surrounded by Boltzmann distributed charge carriers, introduced originally by Debye and H\"uckel \cite{DebyHueckel} and generalized to non Coulombic potentials in Ref.~\cite{Dixit}. While it was shown that the resulting parametrization of Re[V] can reproduce the lattice data quite well, it required the introduction of a second temperature dependent fit parameter which complicates the interpretation of the screening in terms of a unique screening mass. Since this classical approximation is unable to accommodate an imaginary part of the potential, no statement about Im[V] was made. On the other hand an interesting approach was proposed in Ref.~\cite{Thakur:2013nia}, which attempts to describe the screening of the potential in terms of actual in-medium field theory. The authors make the assumption that the in-medium potential arises from the vacuum potential by multiplying it with a field-theory determined complex permittivity in momentum space. Using the perturbative permittivity calculated in the HTL approximation it was possible to reproduce the known real- and imaginary part of the potential in HTL. Unfortunately applying the permittivity to the linearly rising potential lead to unphysical results. The real-part does not decay exponentially but retains an $\sim 1/r$ behavior, which does not describe the lattice potential and hints at the fact that the screening of the vacuum potential is not captured self-consistently. Even more severely, the resulting imaginary part, which from physical considerations must go to a constant at large distance, namely twice the Landau damping of a single quark, diverges logarithmically.

Our study combines the strength of these two approaches bringing together the concept of a generalized Gauss law from Ref.~\cite{Dixit} with the characterization of in-medium effects through the perturbative HTL permittivity. The use of the Gauss law, a non-local concept, leads to a self-consistent description of both screening and damping effects evading the unphysical behavior of Ref.~\cite{Thakur:2013nia}.

\section{The static in-medium inter-quark potential} \label{pot}

The static potential acting between a color singlet pair of a quark and antiquark immersed in a thermal medium of light quarks and gluons constitutes the basis for our gauge invariant QCD screening mass. The definition of the heavy quark potential at finite temperature itself has been a long standing problem in thermal field theory. Using an effective field theory description \cite{Barchielli:1986zs}, based on the separation of scales between the mass of the heavy quark and its typical momenta, it had been worked out how to define the potential in vacuum from a dynamical QCD observable, the real-time Wilson loop $W(t,r)$. With the maturation of modern effective field theories \cite{Brambilla:2004jw}, such as NRQCD and pNRQCD it became possible to take into account the presence of the additional scale represented by temperature and to systematically extend the validity of the potential definition to finite temperature \cite{Brambilla:2008cx}. It reads
\beq
V(r)=\lim_{t\to\infty} \frac{i\partial_t W(t,r)}{W(t,r)}\label{Eq:VRealTimeDef}.
\eeq
This potential was first calculated in hard thermal loop (HTL) resummed perturbation theory \cite{Laine:2007qy} and was found to be complex valued:
\beq
V_{\rm HTL}(r)= -\tilde\alpha_s\left[\mD+\frac{e^{-\mD r}}{r}
+iT\phi(\mD r)\right]+\mathcal{O}(g^4)\label{Eq:VHTL},
\eeq
where
\beq
\phi(x)=2 \int_0^\infty dz \frac{z}{(z^2+1)^2}\left(1-\frac{\sin(xz)}{xz}\right)\label{phi}
\eeq
and a factor $C_F$ has been absorbed in the definition of the coupling constant $\tilde\alpha_s=\frac{g_s^2C_F}{4\pi}$ to match the literature on phenomenology. Its real-part indeed shows the typical Debye screened behavior. A non-perturbative, i.e. lattice QCD based, determination of the potential however remained a conceptual and technical challenge, which has only recently been overcome in a satisfactory way \cite{Burnier:2014ssa,Rothkopf:2011db,Burnier:2012az,Burnier:2013fca,Burnier:2013nla}. The central hurdle is related to the fact that lattice simulations are performed in Euclidean time and have no direct access to real time quantities such as $W(r,t)$.

A possible way around this limitation was proposed in Ref.~\cite{Rothkopf:2009pk} with a first attempt at an implementation presented in Ref.~\cite{Rothkopf:2011db}. The underlying idea is to use a spectral decomposition of the Euclidean Wilson loop $W(\tau,r)$ to relate the Euclidean an Minkowski time domain 
\beq
 \nonumber W(\tau,r)=\int d\omega e^{-\omega \tau} \rho(\omega,r)\,
\leftrightarrow\, \int d\omega e^{-i\omega t} \rho(\omega,r)= W(t,r).
\eeq
This equations can be combined with Eq.~\eqref{Eq:VRealTimeDef} to define the potential in terms of the Wilson loop spectral function
\begin{align}
\hspace{-0.2cm}V(r)=\lim_{t\to\infty}\int d\omega\, \omega e^{-i\omega t} \rho(\omega,r)/\int d\omega\, e^{-i\omega t} \rho(\omega,r). \label{Eq:PotSpec}
\end{align}
At that stage the definition requires precise knowledge of the spectrum of the Euclidean Wilson $\rho(\omega,r)$, which can in principle be obtained from an inverse Laplace transform of datapoints $W(\tau_n,r),~n=1..N_\tau$ simulated in lattice QCD. Bayesian inference plays an important role to give meaning to this otherwise ill-posed problem. In practice however carrying out the inverse Laplace transform posed a formidable challenge to standard methods, such as the Maximum Entropy Method \cite{Asakawa:2000tr} or extended MEM \cite{Rothkopf:2012vv}. In fact, in the end it required the development of a novel Bayesian inference method \cite{Burnier:2013nla}.

Even if the Wilson loop spectrum can be determined reliably from the lattice, a second difficulty lies in the infinite time limit in Eq.~(\ref{Eq:PotSpec}). Using only the symmetries of the real-time Wilson loop, it was shown in Ref.~\cite{Burnier:2012az} that the physics of the potential manifests itself in the lowest lying peak in the spectrum, which takes the shape of a skewed Lorentzian
\begin{align}
&\rho\notag\propto\frac{|{\rm Im} V(r)|{\rm cos}[{\rm Re}{\sigma_\infty}(r)]-({\rm Re}V(r)-\omega){\rm sin}[{\rm Re} {\sigma_\infty}(r)]}{ {\rm Im} V(r)^2+ ({\rm Re} V(r)-\omega)^2}\\ \notag&+{c_0}(r)+{c_1}(r)({\rm Re} V(r)-\omega)+{c_2}(r)({\rm Re} V(r)-\omega)^2\ldots\,.
\end{align}
Inserting this functional form of the spectrum into the definition \eqref{Eq:PotSpec} it was confirmed that the position and the width of this peak encode the values of the real- and imaginary part of the potential \cite{Burnier:2013wma}. 

To put this extraction strategy into practice we generated anisotropic $\xi_b=3.5$ quenched QCD configurations at $\beta=7$ with temporal extend $N_\tau=24\ldots96$, i.e. spanning $T=839\ldots210$MeV ($T_c\approx271$MeV), and extracted both the real and imaginary part of the potential as reported on in Ref.~\cite{Burnier:2014ssa}. For the current work we added an additional set of $N_{\rm conf}=900$ low temperature configurations at $N_\tau=192$, i.e. $T=105$MeV, that will be used in the determination of the Debye mass.

\section{An analytic parametrization of the heavy-quark potential}\label{fit_fct}

The starting point of our derivation is the generalized Gauss law introduced in Ref.~\cite{Dixit} for an electric field the form $\vE=q r^{a-1} \rh$
\beq
\vec\nabla \left(\frac{\vE}{r^{a+1}}\right)=   4\pi \delta(\vr) \label{Eq:GenGauss}.
\eeq
Using the relation $-\vec\nabla V(\mathbf{r})=\vE(\mathbf{r})$ Eq.~\ref{Eq:GenGauss} reduces to the well known expression for the Coulomb potential for $a=-1, q=\tilde\alpha_s, [\tilde\alpha_s]=1$, while a linearly rising potential corresponds to $a=1, q=\sigma, [\sigma]={\rm GeV}^2$.

Let us first have a look at the original argument by Debye and H\"uckel. In their work the above equation is modified through the presence of a background charge density $\langle \rho(\vec r) \rangle$
\beq 
\vec\nabla \left(\frac{\vE}{r^{a+1}}\right)=4\pi\big( \delta(\vr)+\langle \rho(\vec r)\rangle\big), \label{Eq:nablaE}
\eeq
which represents a Boltzmann's distribution for the charge carriers at temperature $T=1/\beta$, 
\beq
\langle \rho(\vec r)\rangle=q\left(n_0 e^{- \beta V(\vr)}-n_0\right)-q\left(n_0 e^{\beta V(\vr)}-n_0\right)\label{Eq:OldDebHueck},
\eeq
where in our context the first term stands for particles and the second for antiparticles. $n_0$ is the charge density in the absence of the test charge. If the resulting in-medium potential is weak, we can expand the exponential in Eq.~\ref{Eq:OldDebHueck} as
\beq
\langle\rho(\vec r)\rangle=-2q\beta n_0 V(\vr).
\eeq
When plugged into Eq.~(\ref{Eq:nablaE}), we obtain \cite{Dixit}:
\beq
-\frac{1}{r^{a+1}}\nabla^2 V(r)+\frac{1+a}{r^{a+2}}\nabla V(r)+A V(r)=4 \pi q \delta(\vr),\label{Eq:phi}
\eeq
where $A=8 \pi q n_0 \beta$. The appearance of the term with prefactor $A$ is a manifestation of the linear-response
character of this approximation. For the Coulombic part of the potential ($a=-1, q=\tilde\alpha_s$) we have
\beq
-\nabla^2 V_C(r)+A_C V_C=4 \pi \tilde\alpha_s \delta(\vr).\label{Eq:Coulomb(r)}
\eeq
On the other hand for the string case ($a=1,q=\sigma$) one finds
\beq
-\frac{1}{r^{2}}\frac{d^2 V_s(r)}{dr^2}+A_s V_s(r)=4 \pi \sigma \delta(\vr),\label{Eq:String(r)}
\eeq
where $A_s=8 \pi \sigma n_0 \beta$. If $n_0$ is indeed the unmodified charge density it has to take the same vale for both cases, which will allow us to relate $A_C$ and $A_s$ in the following. 

Now let us return to the generalized Gauss law of Eq.~\ref{Eq:GenGauss} for a Coulomb charge $(a=-1)$ written in momentum space
\beq
p^2 V_C(\vp)=4\pi\tilde\alpha_s.
\eeq
In the language of field theory, the effects of the medium is to dress the point charge with a cloud of charge carries. It amounts to washing out the coordinate space delta function on the RHS with an in-medium permittivity $\epsilon(\vp,m_D)$, which encodes the full temperature dependence of the problem:
\beq
p^2 V_C(\vp)=4\pi\frac{\tilde\alpha_s}{\eps(\vp,m_D)}\label{Eq:phi(p)}.
\eeq
Similar to Ref.\cite{Thakur:2013nia}, we use the perturbative HTL expression
\beq
\eps^{-1}(\vp,m_D)=\frac{p^2}{p^2+m_D^2}-i\pi T \frac{p\, m_D^2}{(p^2+m_D^2)^2},
\eeq
henceforth assuming a medium of weakly coupled quarks and gluons, in which our test charge is immersed. Inserting this formula into equation (\ref{Eq:phi(p)}) and multiplying by $\frac{p^2+m_D^2}{p^2}$, we obtain:
\beq
p^2  V_C(\vp)+m_D^2 V_C(\vp)=4\pi\tilde\alpha_s \Big( 1 -i\pi T\frac{ m_D^2}{p(p^2+m_D^2)} \Big).\label{Eq:Coulomb(p)}
\eeq
The inverse Fourier transform of the real part of Eq.~\eqref{Eq:Coulomb(p)} exactly reproduces the linear-response expression of \eqref{Eq:Coulomb(r)}, allowing us to identify $A_C=m_D^2$ and in turn gives an expression for the charge density
\beq
n_0=\frac{m_D^2 T}{8 \pi \tilde\alpha_s}.\label{Eq:n_0}
\eeq
The imaginary part arising on the RHS of Eq.~\eqref{Eq:Coulomb(p)} can also be Fourier transformed to coordinate space, which completes our generalized formula for the in-medium modification of a Coulombic test charge
\beq
-\nabla^2 V_C(r)+m_D^2V_C(r)=\tilde\alpha_s \Big(  4 \pi  \delta(\vr)-  iT m_D^2 g(m_Dr)\Big),\label{Eq:Coulomb(r)withIm}
\eeq
with 
\beq
g(x)=2\int_0^\infty dp  \frac{\sin(p x)}{p x}\frac{p}{p^2+1}.
\eeq
The solution to Eq.~\eqref{Eq:Coulomb(r)withIm} with the physical boundary condition $\left.{\rm Re}V_C(r)\right|_{r=\infty}=0$, $\left.{\rm Im}V_C(r)\right|_{r=0}=0$ and $\left.\partial_r {\rm Im}V_C(r)\right|_{r=\infty}=0$ coincides with the HTL potential (\ref{Eq:VHTL}) obtained by Laine et al.~\cite{Laine:2007qy}
\begin{align}
{\rm Re}V_C(r)=-\frac{\tilde\alpha_s}{r}\left( \frac{e^{-m_D r}}{r} +  m_Dr  \right),~~
{\rm Im}V_C(r)= -\tilde\alpha_s\,T\,\phi(x) ,\label{Eq:VCHTL} 
\end{align}
where $\phi(x)$ has been defined in Eq.~(\ref{phi}). 

Now we turn to the string-like, i.e. linearly rising part of the test-charge potential. Note that in contrast to the Coulomb case, for $a=1$ we cannot transform Eq.\eqref{Eq:GenGauss} into a simple relation for the Fourier space potential akin to Eq.\eqref{Eq:phi(p)}. Instead, as a first step, we return to Eq.~(\ref{Eq:String(r)}) to obtain an explicit expression for the coefficient $A_s$. Taking the charge density $n_0$ to be the same, irrespective of the test charge being Coulombic or string-like one then obtains together with Eq.~(\ref{Eq:n_0}) 
\beq
A_s=\mu^4=m_D^2\frac{\sigma}{\tilde\alpha_s}. \label{Eq:mu}
\eeq
In the context of the linear-response pitcture this tells us that the strength of the in-medium modification of the linear potential is controlled by the parameter $\mu$ and not $m_D$ itself. To arrive at the defining equation for the medium modified string-potential, we assume the validity of the linear response approximation if the charge density on the RHS of Eq.~\eqref{Eq:String(r)} is replaced by the medium modified charge density obtained from HTL in Eq.~\eqref{Eq:Coulomb(r)withIm}
\beq
-\frac{1}{r^{2}}\frac{d^2 V(r)}{dr^2}+ \mu^4 V(r)=\sigma \Big( 4 \pi \delta(\vr)- i T m_D^2 g(m_D r)\Big).\label{Eq:String(r)withIm}
\eeq
Note that these expressions differ significantly from the ones used in \cite{Digal:2005ht}, where $A_s$ was chosen on purely dimensional grounds to be $m_D^4$. As we will see in the next section, our choice motivated by the charge density allows one to adequately fit the lattice data as is and thus makes the introduction of any other fitting parameter unnecessary. 

Solving for the real part of Eq.~\eqref{Eq:String(r)withIm} with the same physical boundary conditions as in the Coulomb case we find
\begin{align}
{\rm Re}V_s(r)&=\frac{\Gamma[\frac{1}{4}]}{2^{\frac{3}{4}}\sqrt{\pi}}\frac{\sigma}{\mu} D_{-\frac{1}{2}}\big(\sqrt{2}\mu r\big)+ \frac{\Gamma[\frac{1}{4}]}{2\Gamma[\frac{3}{4}]} \frac{\sigma}{\mu},
\end{align}
which besides the definition of $\mu$ also differs in a factor $\sqrt{2}$ in the argument of the parabolic cylinder function $D_\nu(x)$ compared to Ref.\cite{Digal:2005ht}. The imaginary part of the in-medium modified string potential can be written in a closed form as a Wronskian solution:
\begin{eqnarray}
\Im V_s(r)&=&-i\frac{\sigma m_D^2 T}{\mu^4} \psi(\mu r)=-i\tilde\alpha_s T \psi(\mu r)\label{Eq:DebHueck},
\end{eqnarray}
with
\begin{eqnarray} 
\notag\psi(x)&=&D_{-1/2}(\sqrt{2}x)\int_0^x dy\, {\rm Re}D_{-1/2}(i\sqrt{2}y)y^2 g(ym_D/\mu)\\\notag&&+{\rm Re}D_{-1/2}(i\sqrt{2}x)\int_x^\infty dy\, D_{-1/2}(\sqrt{2}y)y^2 g(ym_D/\mu)\\&&-D_{-1/2}(0)\int_0^\infty dy\, D_{-1/2}(\sqrt{2}y)y^2 g(ym_D/\mu),
\end{eqnarray}
This completes our derivation of the analytic expression for the real- and imaginary part of the in-medium QCD heavy-quark potential.

Let us have a look at the behavior of the solution just obtained. In the limit of zero temperature, i.e. vanishing $m_D$, we recover the Cornell potential in the real part. As expected, at small distances the Coulombic real part behaves as $1/r$ whereas the string shows a linear rise with $r$. The imaginary part on the other hand rises according to $r^2$ for the Coulombic part and with $r^3$ for the string. I.e. the Coulombic HTL part (\ref{Eq:VCHTL}) dominates at small $r$. At large distances we again find that the Coulombic part dominates the real part and behaves just like the naive Debye screened potential $\exp(-m_D r)/r$. The fact that the string part dies off much more rapidly as $\exp(-m_D^2 r^2/2)$ is the reason why we can actually identify the parameter $m_D$ with the Debye mass, when fitted to the functional form of the lattice potential. At asymptotically large distances both the Coulombic and string imaginary part saturate to a constant as required.

\section{Determining the Debye Mass through the complex in-medium heavy-quark potential from lattice QCD}

Our goal is to use the derived analytic expression for the in-medium potential to extract the Debye mass from the static inter-quark potential recently measured in lattice QCD. In this work we focus on the case of a purely gluonic medium, for which both ${\rm Re}V$, as well as ${\rm Im}V$ have been determined at various temperatures in the phenomenologically relevant region around the deconfinement transition. Since only a single fit parameter $m_D$ is required, we will carry out fits solely to the real-part of the potential, so that the agreement or disagreement of the corresponding imaginary part can serve as a crosscheck of our approach.

\begin{figure}[t!]
\hspace{-0.5cm} \includegraphics[scale=0.3]{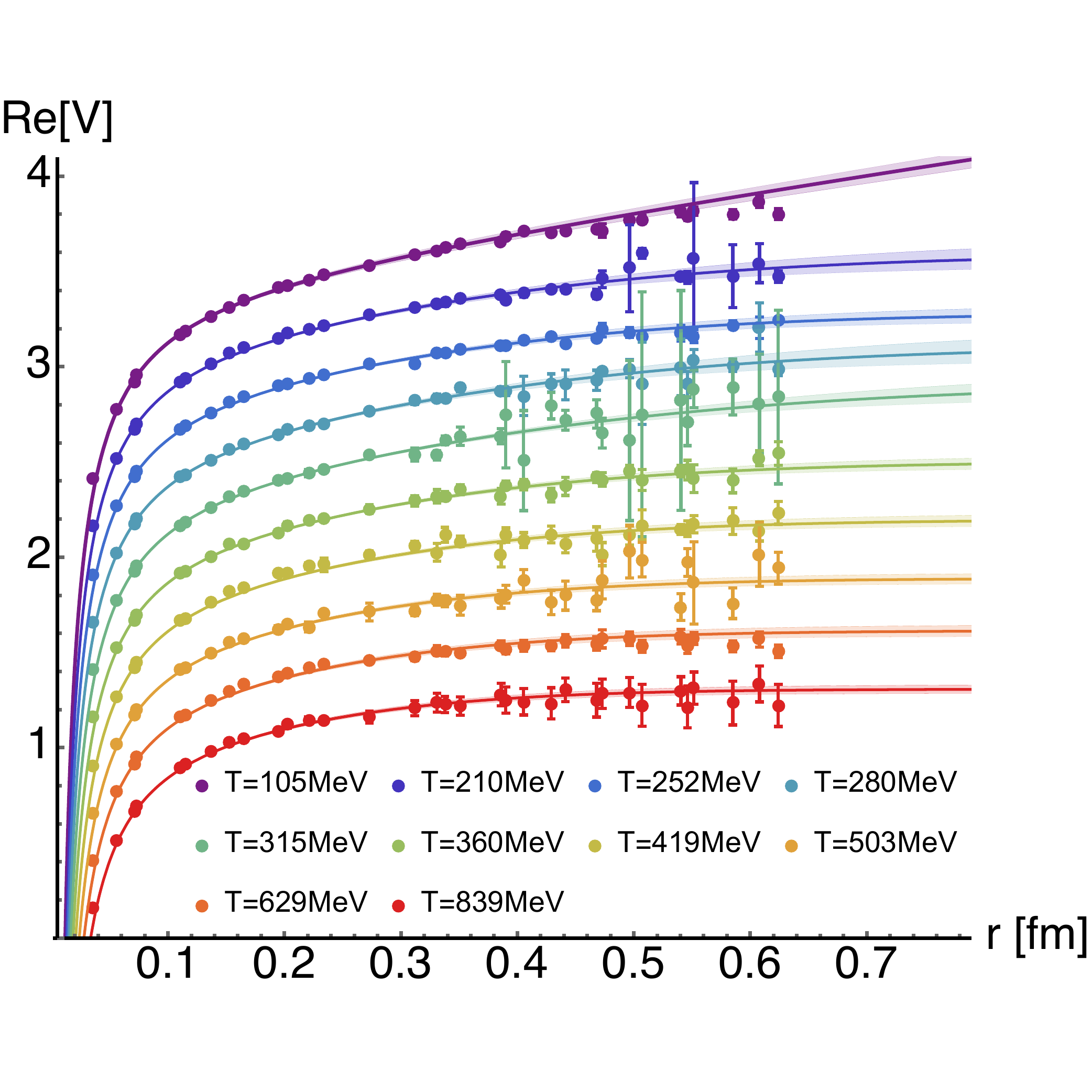}\hspace{0.2cm}\includegraphics[scale=0.3]{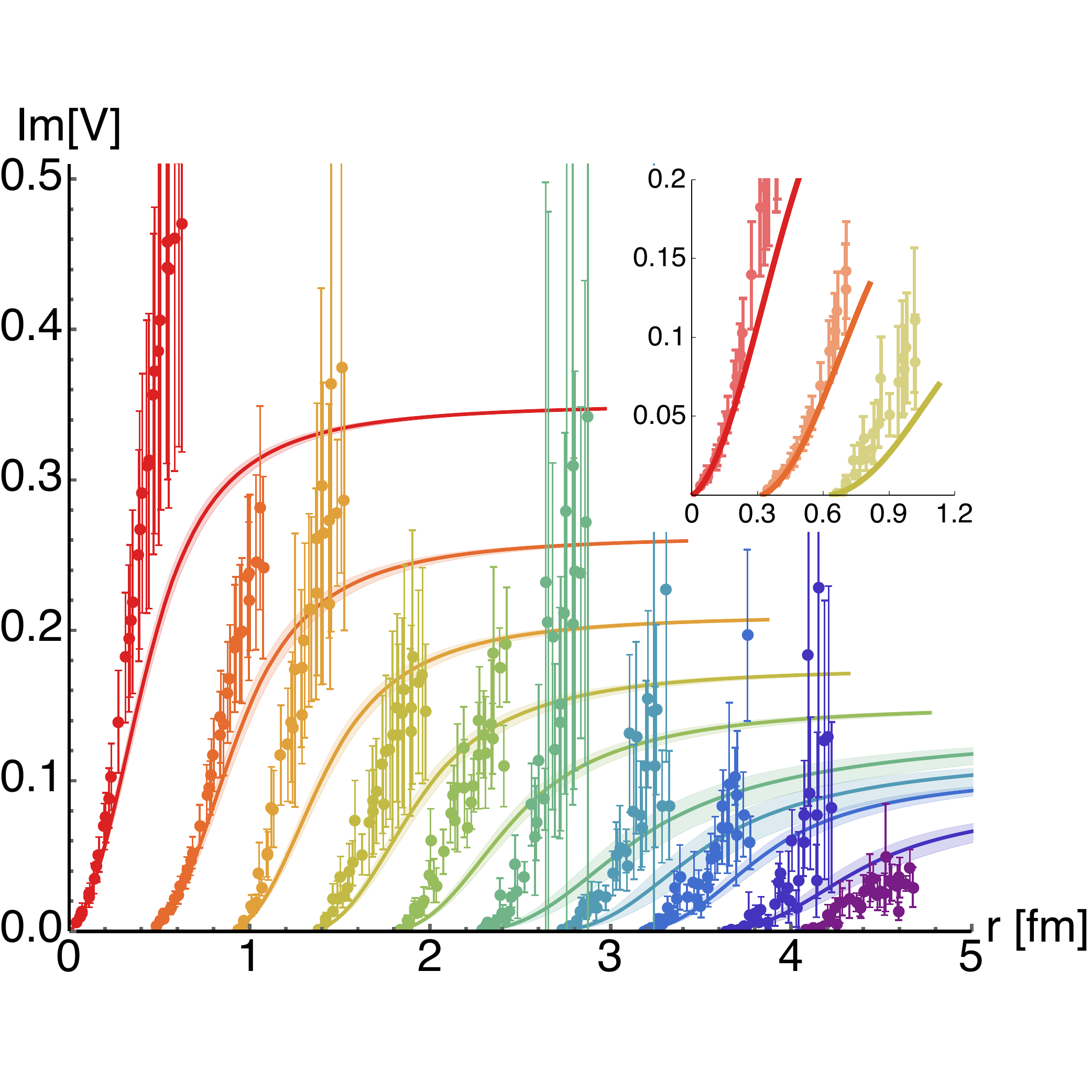}
\vspace{-1cm}
\caption{ (left) One parameter fit (solid lines) to the real part of the potential (data points) obtained in quenched QCD. (right) Imaginary part of the potential in quenched QCD (data points) and the values obtained from our analytic expression using the Debye mass fitted in ${\rm Re}V$.} 
\label{fig:fit_reim} 
\end{figure}

We assume the values of the strong coupling $\tilde\alpha_s$ and string tension $\sigma$ not to vary with temperature $T$, as they characterize the properties of the test charge to be inserted in the medium. Their values hence have to be determined in vacuum. In the absence of a true $T=0$ lattice measurement, we use the newly generated lattice ensemble at $T=105$MeV instead and fit the small to intermediate $r$ region of ${\rm Re}V$. There lattice cutoff artifacts are expected to be small, the lattice determination of the real-part is most reliable and the effects of the small but finite temperature are not yet significant. The particular nature of the lattice normalization of the potential introduces a constant shift $c$, which we also determine
\beq
\tilde\alpha_s=0.206\pm0.011, \quad \sigma=0.174\pm0.011 {\rm GeV}^2, \quad c=2.60\pm0.023 {\rm GeV}.
\eeq
Varying the fitting range up to a maximum of six steps at the upper and lower end of the fitting interval yields the error estimates shown. The only remaining free parameter at finite temperature then is the Debye mass $m_D$. As can be seen by the agreement of the solid lines and data points in the left panel of Fig.\ref{fig:fit_reim}, its tuning alone allows us to achieve an excellent fit of the real-part of the potential at all temperatures, both qualitatively and quantitatively. To account for the propagation of the error on the low temperature parameters, besides changing the fitting range on the finite $T$ potential, we also use in each range different combinations of the values for $\tilde\alpha_s,\sigma$ and $c$ according to the uncertainties from the fits at $T\simeq0$. The values we obtain for the Debye mass together with their error estimates are given in Tab.\ref{Tab:DebyeMass}.

\begin{table}[b!]
\small
\hspace{-2.5cm}\begin{tabularx}{17cm}{ | c | X | X | X | X | X | X | X | X | X | X |}
\hline
	SU(3): $N_\tau$ & 24 & 32 & 40 & 48 & 56 & 64 & 72 & 80 & 96 & 192\\ \hline
	$T$[MeV] & 839 & 629 & 503 & 419 & 360 & 315 & 280 & 252 & 210 & 105\\ \hline
	$N_{\rm meas}$ & 3270 & 2030 & 1940 & 1110 & 1410 & 1520 & 860 & 1190 & 1800 & 900\\ \hline
	$m_D[MeV]$ & $ 852\pm60$ & $709\pm70$ & $654\pm62$ & $537\pm59$ & $444\pm52$ & $264\pm62$ & $309\pm86$ & $401\pm62$ & $328\pm78$& 0\\ \hline	
\end{tabularx}\caption{Quenched lattice parameters ($\beta=7,\xi_b=3.5, N_s=32$) and values of the extracted Debye mass}\label{Tab:DebyeMass}
\end{table}

Note that while we have determined $m_D$ solely from an inspection of the real-part, the resulting values for ${\rm Im}V$ also agree reasonably well with the lattice data (Fig.\ref{fig:fit_reim} right). At high temperatures and small distances, where the lattice reconstruction of the potential is most reliable, we even find quantitative agreement with the analytical form within statistical errors. At larger distances we expect that the lattice data-points are indeed larger than the actual values of ${\rm Im}V$, as the underlying extraction from spectral widths leads to unphysically large values due to a diminishing signal to noise ratio.

Close to $T_C$ at $T=315$MeV our analytic postdiction of ${\rm Im}V$ appears to lie rather far away from the lattice data. The reason for this, we believe, is twofold. On the one hand the extracted values for ${\rm Im}V$ at this temperature are rather imprecise, with the statistical errorbars still sizeable. On the other hand the determination of $m_D$ from ${\rm Re}V$ also seems to lack accuracy. Indeed if we inspect the corresponding fit curve on the left of Fig.\ref{fig:fit_reim}, we can see already by eye that is too steep compared to those at neighboring temperatures. Using a higher statistics estimate for $m_D$ from ${\Re V}$ at temperatures around the phase transition in the future, will most certainly yield closer agreement between the analytic expression and the actual lattice imaginary part.

\section{Debye mass and its match to HTL}\label{mD}

The temperature dependence of the extracted Debye masses can be compared to predictions from resummed HTL perturbation theory. According to Ref.~\cite{Arnold:1995bh} the Debye mass at leading log order can be written as:
\begin{eqnarray}
m_D&=&T g(\mu T)\sqrt{\frac{N_c}{3}+\frac{N_f}{6}} \notag
	+\frac{N_c T g(\mu  T)^2}{4 \pi } \log
   \left(\frac{\sqrt{\frac{N_c}{3}+\frac{N_f}{6}}}{g(\mu  T)}\right)
   \\&&+c \,T g(\mu  T)^2+d\, T g(\mu  T)^3, \label{Eq:mD}
\end{eqnarray}
where $\mu$ denotes the renormalization scale, $g$ the running QCD coupling, and $c$ and $d$ are constants that represent non-perturbative contributions which need to be determined from a fit to the data.
\begin{figure}[t!]
\hspace{-0.8cm} \includegraphics[scale=0.26]{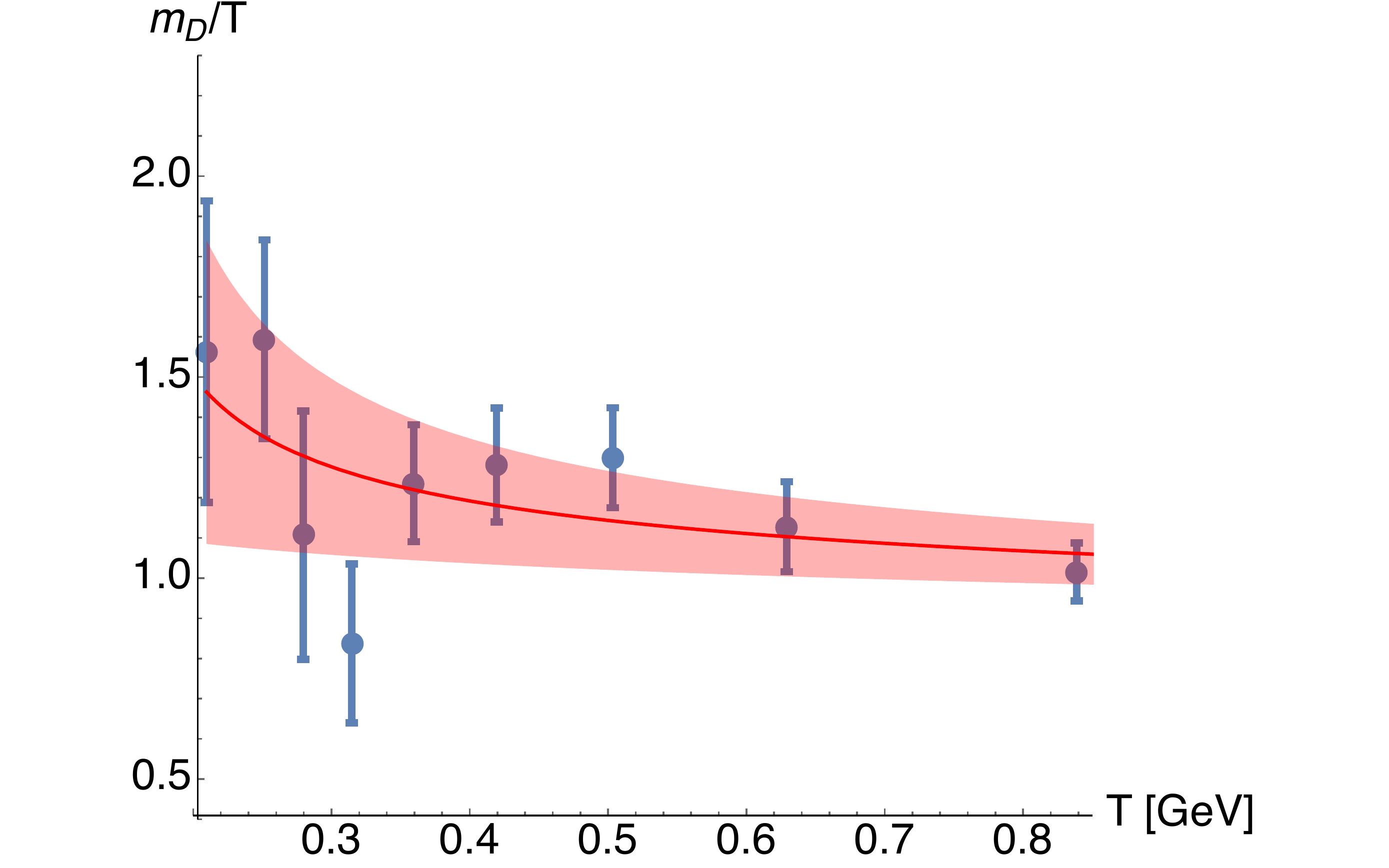}\hspace{-1.2cm}\includegraphics[scale=0.26]{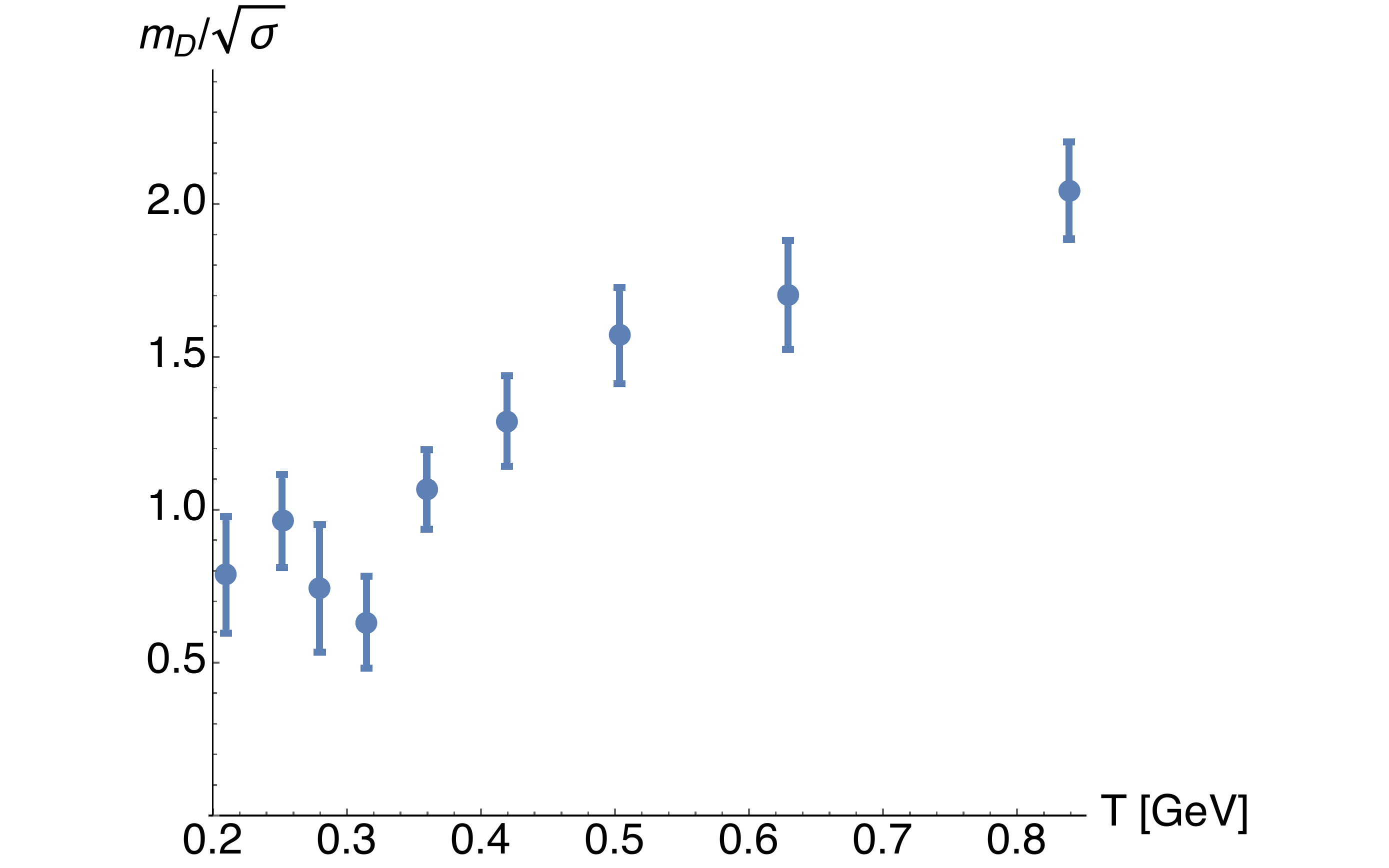}
\vspace{-0.2cm}
\caption{(left) Debye masses (blue) and HTL fit (red) of the temperature dependence obtained on quenched lattices. Error bars includes both the statistical error, as well as those from changing the fitting range and propagation of the error of $\tilde\alpha_s$ and $\sigma$ determined at low temperature. (right) The dimensionless ratio of $m_D/\sqrt{\sigma}$ for use in phenomenological modeling.} 
\label{fig:mD_qu} 
\end{figure}
As there exists a clear degeneracy between the variation of the renormalization scale and the variation of the higher order contributions, parametrized by $c$ and $d$, we choose to fix $\mu$ according to the usual convention $\mu=\pi T$ and fit $c,d$ from the obtained Debye masses in the previous section. 

For the running of the coupling $g(\mu)$ we utilize the four loop result of Ref.~\cite{vanRitbergen:1997va} setting $\Lambda_{QCD} = 0.216$, appropriate for quenched QCD. The fit yields $c=-0.40\pm0.06$ and $d=0.21\pm0.06$, which is shown in Fig.~\ref{fig:mD_qu}. Note that the values obtained for $c$ and $d$ are small, which implies quite good agreement between hard thermal loop perturbation theory and the lattice extraction, even at the low temperatures probed here. In fact, one could argue that for a HTL fit, only the high temperature points should be used, for which a perturbative expansion is expected to apply. We see however that including the lower temperatures close to $T_c$ does not change the determination of the parameters $c,~d$ in an significant way. 

In previous studies in quenched QCD, Debye masses were e.g. obtained by fitting not the proper heavy-quark potential but the color-singlet free-energies with a simple Coulombic Debye-screened form \cite{Maezawa:2007fc}. Comparing, we find that the values obtained here lie consistently lower than these previous estimates.

We would like to note that even though the lattices used to determine the heavy-quark potential deployed in this study are quite finely spaced, no continuum extrapolation has been carried out. Therefore it is doubtful whether the values of the Debye mass shown on the left of Fig.\ref{fig:mD_qu} together with the values for $\tilde\alpha_s$ and $\sigma$ can be directly used in phenomenological models in the continuum. As a possible remedy we follow Ref.~\cite{Digal:2005ht} in providing the ratio of $m_D/\sqrt{\sigma}$ on the right of Fig.\ref{fig:mD_qu}, in which some of the systematic uncertainties arising from a finite lattice spacing might be expected to cancel. 

\section{Conclusion}

Based on a combination of the generalized Gauss law, introduced in \cite{Dixit} and the in-medium modification of its point charge distribution by a weakly interacting medium of light quarks and gluons, described by the HTL permittivity, we derived an analytic expression for the real- and imaginary part of the static inter-quark potential at finite temperature. After fixing the strong coupling and string tension at low temperature, we are able to qualitatively and quantitatively reproduce the real-part of the potential measured in lattice QCD by fitting a single temperature dependent parameter $m_D(T)$, which is proposed as gauge invariant screening mass in QCD. The temperature dependence of $m_D$ we obtained in a purely gluonic medium agrees well with the predictions of HTL perturbation theory. Using the fitted values for $m_D$ we furthermore find that a quite successful postdiction of ${\rm Im}V$ at high temperatures is possible. Agreement with ${\rm Im}V$ at smaller temperatures, currently hampered by uncertainties in the fit of $m_D$ to the real-part, should improve once higher statistics has been collected.

We hope that phenomenological modeling will benefit from the derivation of a well motivated and lattice data validated parametrization of both ${\Re}V$ and ${\Im}V$. In addition our study opens up the possibility to extract the imaginary part of the potential in full QCD simulations, in which up to date only the real part has been determined in a reliable fashion.

\section*{Acknowledgments}

The authors thank O. Kaczmarek, P. Petreczky and R. Pisarski for stimulating discussions. YB is supported by SNF grant PZ00P2-142524.

\end{document}